\documentclass{article}
\usepackage[a4paper, left=2cm, right=2cm]{geometry}
\usepackage{graphicx} 
\usepackage{amsmath}
\usepackage{amsfonts}
\usepackage{amssymb}
\usepackage{xcolor}
\usepackage{algorithm}
\usepackage{algpseudocode}
\usepackage{stmaryrd}
\usepackage{subcaption}
\usepackage[latin1]{inputenc}

\newtheorem{theorem}{Theorem}
\newtheorem{proposition}{Proposition}
\newtheorem{definition}{Definition}

\newtheorem{lemma}{Lemma}
\newtheorem{remark}{Remark}

\newcommand{\interval}[1]{\left[#1\right]}
\newcommand\intsimple[1]{\left\llbracket #1 \right\rrbracket}
\newcommand{\myvec}[2]{%
	\left[
	\begin{array}{c}
		#1\\
		#2
	\end{array}
	\right]
}
\newcommand{\ub}[1]{\bar{#1}}
\newcommand{\lb}[1]{\underline{#1}}
\newcommand{\II}{\mathcal{I}}
\newcommand{\MM}{\mathcal{M}}
\newcommand{\RR}{\mathbb{R}}
\newcommand{\TT}{\mathbb{T}}
\newcommand{\XX}{\mathcal{X}}
\newcommand{\UU}{\mathcal{U}}
\newcommand{\WW}{\mathcal{W}}
\renewcommand{\SS}{\mathcal{S}}
\newcommand{\ZZ}{\mathcal{Z}}
\newcommand{\BB}{\mathcal{B}}
\newcommand{\EE}{\mathcal{E}}
\newcommand{\Ac}{\hat{A}}
\newcommand{\Bc}{\hat{B}}
\newcommand{\AKc}{\hat{A}_K}
\newcommand{\AK}{\mathcal{A}_K}
\newcommand{\MD}{\II_S}
\newcommand{\MDK}{\II_{A_K}}
\newcommand{\Mdelta}{\II_{\Delta}}
\newcommand{\bbox}[1]{\mathbb{B}\left(#1\right)}
\newcommand{\qed}{\hfill$\square$}
\newcommand{\Pk}{\mathbb{P}_k(x(k),\ZZ_{f,k})}

\newcommand{\new}[1]{\textcolor{black}{#1}}

\title{Time-Optimal Model Predictive Control\\ for Linear Systems with Multiplicative Uncertainties}
\author{
	Renato Quartullo$^{1}$, Andrea Garulli$^{1}$, and Mirko Leomanni$^{2}$\thanks{$^{1}$Dipartimento di Ingegneria dell'Informazione e Scienze Matematiche, Universit\`a di Siena, Italy. Email: \texttt{\{quartullo, garulli\}@diism.unisi.it}.}\thanks{$^{2}$Dipartimento di Ingegneria e Scienze, Universit\`a Mercatorum, Italy. Email: \texttt{mirko.leomanni@unimercatorum.it}.}
}
\date{}

\begin{document}
	
	\maketitle
	
	\begin{abstract}
		This paper presents a time-optimal Model Predictive Control (MPC) scheme for linear discrete-time systems subject to multiplicative uncertainties represented by interval matrices. To render the uncertainty propagation computationally tractable, the set-valued error system dynamics are approximated using a matrix-zonotope-based bounding operator. Recursive feasibility and finite-time convergence are ensured through an adaptive terminal constraint mechanism. A key advantage of the proposed approach is that all the necessary bounding sets can be computed offline, substantially reducing the online computational burden. The effectiveness of the method is illustrated via a numerical case study on an orbital rendezvous maneuver between two satellites.
	\end{abstract}
	
	\section{Introduction}
	
	Model Predictive Control (MPC) is a popular approach that has proved to be successful in handling a wide variety of control problems and for which a well-established theory is available \cite{rawlings2017model,borrelli2017predictive}.
Among the many different formulations of MPC-based control schemes, a special setting is that in which the length of the prediction horizon, instead of being fixed a priori, is included among the variables of the optimization problem. This is particularly useful in applications in which the control task must be performed within a given (or minimum) amount of time. Examples of such MPC schemes can be found in mobile robotics for point-to-point motion design \cite{verscheure2009time,ardakani2018model}, helicopter landing maneuvers \cite{greer2020shrinking}, spacecraft rendezvous missions \cite{leomanni2022variable} and optimal train operations \cite{farooqi2020shrinking}.

In minimum-time (or time-optimal) MPC, the cost function to be minimized in the optimization problem is the length of the prediction horizon itself. Several solutions to this problem can be found in the literature. In \cite{van2011model}, the horizon length minimization is pursed by adopting a bilevel optimization algorithm. An approach exploiting time-elastic bands is proposed in \cite{rosmann2015timed}. Time optimality is achieved indirectly in \cite{verschueren2017stabilizing} by introducing an exponential weight in the cost function. Conditions for closed-loop Lyapunov stability of minimum-time MPC are derived in \cite{sutherland2019closed}. 

The works cited so far assume that the system is not affected by uncertainties. However, there is a vast literature on how different type of uncertainty sources can be handled within the MPC framework. The presence of bounded additive disturbances has been treated extensively, leading to solutions based on min-max control \cite{bemporad2003min,buerger2014active} or on the construction of tubes bounding the sequence of predicted states and inputs (see, e.g., \cite{lee1999constrained,chisci2002feasibility,mayne2005robust,rakovic2012homothetic}).
The case of multiplicative uncertainties affecting the system dynamics is more challenging and it is still the subject of ongoing research.  The approaches proposed in the literature to tackle this problem range from ellipsoidal invariant sets \cite{kothare1996robust} to {polytopic or zonotopic tubes \cite{fleming2014robust,munoz2015robust,lu2019robust,russo2023tube}} and more recent optimization-based constraint tightening techniques \cite{BORRELLI22}. All these methods essentially seek a trade-off between the conservatism of the uncertainty propagation and the computational burden of the resulting optimization problem. See \cite[Chapter 5]{CANNONbook} for a detailed discussion of the different approaches.
In the context of time-optimal MPC, a solution accounting for the presence of bounded additive disturbances is presented in \cite{quartullo2024robust}. To the best of our knowledge, time-optimal MPC for systems with multiplicative uncertainties has not been addressed so far: this is the aim of the present work.

The main contribution of this paper is a time-optimal MPC scheme for linear discrete-time systems affected by model uncertainties, described in terms of interval matrices. In order to make uncertainty propagation computationally tractable, the uncertain impulse response of the system is approximated by using a bounding operator based on matrix zonotopes. Recursive feasibility and finite-time convergence of the control scheme are achieved by adopting an adaptive terminal constraint mechanism, similar to the one proposed in \cite{quartullo2025robust} for variable-horizon MPC. The advantage of the proposed technique for uncertainty propagation is that all the necessary bounding sets can be computed offline. This significantly reduces the computational burden of the online optimization problem, thus allowing to treat systems of nontrivial dimension. This is demonstrated on a numerical example concerning an orbital rendezvous maneuver between two satellites.

The rest of the paper is organized as follows. Section~\ref{sec:notation} introduces the notation and preliminary concepts required for bounding the involved uncertainty sets. The time-optimal control problem is formulated in Section~\ref{sec:PF}, while the proposed control strategy and its properties are presented in Section~\ref{sec:TOR}. Numerical results for an orbital rendezvous example are presented in Section~\ref{sec:simulations}. Finally, conclusions are drawn in Section~\ref{sec:conclusions}.
	
	\section{Notation and preliminaries}\label{sec:notation}
	This section provides the notation, definitions, and fundamental properties of matrix sets used in the subsequent derivations.

	\subsection{Notation}
	The symbol $\oplus$ represents the \emph{Minkowski sum}. In particular, given two sets $\mathcal{A}$, $\mathcal{B}$, $\mathcal{A}\oplus\mathcal{B} = \{a+b|\,\,a\in\mathcal{A},b\in\mathcal{B}\}$. The \emph{exact set multiplication} is $\mathcal{A}\mathcal{B} = \{ab|\,\,a\in\mathcal{A},b\in\mathcal{B}\}$. Similarly, $\mathcal{A}^j = \{a^j|\,\,a\in\mathcal{A}\}$.
	
	Let $\II=\interval{\lb{I},\,\ub{I}} = \{M\in\RR^{p\times q}|\,\, \lb{I}\leq M \leq \ub{I}\}$ be an \emph{interval matrix}, with 
	$\lb{I},\ub{I}\in\RR^{p\times q}$ and the operator $\preceq$ denotes element-wise inequality. The interval matrix $\II$ can
	also be written as $\II = C \oplus 
	\intsimple{\Delta}$, where $C= (\ub{I} + \lb{I})/2$ is the \emph{center} matrix and $
	\intsimple{\Delta} =  \interval{-\Delta,\Delta}$ is the 
	symmetric part of $\II$, with $ \Delta= (\ub{I} - \lb{I})/2$ representing the \emph{radius}.
	A \emph{matrix zonotope} is defined as $$
	\MM = \Big\{ M \in \mathbb{R}^{p \times q} \;\Big|\; 
	M \!=\! M_C + \sum_{i=1}^{e} G^{(i)} \beta_i,\; \|\beta_i\|_{\infty} \leq 1 \Big\},$$
	where $M_C$ is the center, and 
	$G = \big\{ G^{(1)},\, G^{(2)},\, \ldots,\, G^{(e)} \big\}$
	are the \emph{generators}. 
	This matrix zonotope is shortly referred to as $\MM = \langle M_C;\,G^{(1)},\, G^{(2)},\, \ldots,\, G^{(e)}\rangle$. If $q = 1$, we shall simply refer to $\MM$ as a \emph{zonotope}.
	\begin{definition}\label{def:ewdec}
		Given a matrix $ M \in \mathbb{R}^{l \times q} $, we define the \emph{entrywise decomposition operator} as the mapping $\mathcal{E}(M) = \left\{ M^{(k)} \right\}_{k=1}^{lq}$, where each $ M^{(k)} \in \mathbb{R}^{l \times q} $ is a matrix such that
		$$
		M^{(k)}_{ij} =
		\begin{cases}
			M_{ij}, & \text{if } k = (i - 1)l + j, \\
			0, & \text{otherwise}.
		\end{cases}
		$$
	\end{definition}
	{In words, $ \mathcal{E}(M) $ yields a sequence of matrices, each containing exactly one entry of $ M $ in its original position and zeros elsewhere.
		Clearly, $\sum_{k=1}^{lq} M^{(k)} = M$.
		Moreover, the following property holds.
		\begin{proposition}\label{prop:AdecM}
			Let $A\in\mathbb{R}^{n \times l}$ and $M \in \mathbb{R}^{l \times q}$, with $M \succeq 0$. Then, 
			\begin{equation}
				\label{eq:AdecM}
				\sum_{k=1}^{lq} |A M^{(k)} | =|A| M.
			\end{equation}
		\end{proposition}
		\emph{Proof:} From Definition \ref{def:ewdec}, $|A M^{(k)} |= [0 \ldots 0~ |A_i|M_{ij}~ 0 \ldots 0]$, where $A_i$ denotes the $i$th column of A. Then,
		$$
		\sum_{k=1}^{lq} |A M^{(k)} | =\sum_{i=1}^{l}\sum_{j=1}^{q} |A_i| M_{ij}
		$$
		which gives \eqref{eq:AdecM}.\qed
		\\
		In the paper, the shorthand notation $A\mathcal{E}(M)$ is adopted to denote the set of matrices $AM^{(1)}, AM^{(2)}, \ldots, AM^{(lq)}$.}
	
	\subsection{Properties of interval matrices and matrix zonotopes}
	In the following, we list some definitions and properties regarding interval matrices and matrix zonotopes.
	
	Given two interval matrices $\II_1 = C_1 \oplus  \intsimple{\Delta_1}$, $\II_2 = C_2 \oplus  \intsimple{\Delta_2}$, their sum is the interval matrix
	\begin{equation}\label{eq:intsum}
		\II_1 \oplus \II_2 = C_1+C_2 \oplus \intsimple{\Delta_1+\Delta_2}.
	\end{equation}
	The interval matrix product $\II_1 * \II_2$ is defined as
	\begin{equation}\label{eq:interval_product}
		\II_1 * \II_2 
		= {C_1} {C_2} 
		\oplus 
		\intsimple{
			|{C_1}| \Delta_2 
			+ \Delta_1 |{C_2}| 
			+ \Delta_1 \Delta_2
		}.
	\end{equation}
	This interval matrix product provides an over-approximation of the exact set product, that is \cite{rump2012fast}
	\begin{equation}
		\label{eq:intprod}
		\II_1 \II_2 \subseteq \II_1 * \II_2.
	\end{equation}
	By using \eqref{eq:intprod},
	the product of an interval matrix $\II = C \oplus \intsimple{\Delta}$ and a matrix $M$ can be over-approximated as
	\begin{align}
		& \II M = C M \oplus \intsimple{\Delta}M \;\subseteq\; C M \oplus \intsimple{\Delta |M|} \label{eq:intxM}    \\    
		& M \II  = M C \oplus M \intsimple{\Delta} \;\subseteq\; M C \oplus \intsimple{ |M| \Delta} .\label{eq:intxMleft}        
	\end{align}
	
	Let 
	$$
	\MM_1 = \langle M_{C,1};\, G_1^{(1)}, \dots, G_1^{(g_1)} \rangle \subseteq \mathbb{R}^{p \times q},$$
	$$
	\MM_2 = \langle M_{C,2};\, G_2^{(1)}, \dots, G_2^{(g_2)} \rangle \subseteq \mathbb{R}^{p \times q}
	$$
	be two matrix zonotopes.  
	Their Minkowski sum $\MM_1 \oplus \MM_2$ is the matrix zonotope
	$
	\MM_1 \oplus \MM_2 = \langle M_{C,1} + M_{C,2};\, 
	G_1^{(1)}, \dots, G_1^{(g_1)},\, G_2^{(1)}, \dots, G_2^{(g_2)} \rangle.
	$
	\begin{proposition}\label{prop:box}
		The minimum interval matrix containing a matrix zonotope 
		$\MM = \langle M_C;\, G^{(1)}, \ldots, G^{(g)} \rangle$ 
		is given by
		$\II = \bbox{\MM} = M_C \oplus \intsimple{\Delta}$, where $\Delta = \sum_{i=1}^g |G^{(i)}|.
		$
	\end{proposition}

	\begin{definition}\label{def:operatorT}
		Let $\II = C \oplus \intsimple{\Delta} \subset \mathbb{R}^{l \times p}$ be an interval matrix and $\MM = \langle M_C;\, G^{(1)}, \ldots, G^{(e)}\rangle \subset \mathbb{R}^{p \times q}$ be a matrix zonotope. The operator $\TT_{\II}(\MM)$ is defined as
		\begin{equation}
			\label{eq:Topdef}
			\TT_{\II}(\MM) = \langle C M_C;\, CG^{(1)}, \ldots, CG^{(g)},\, F^{(1)}, \ldots, F^{(lq)} \rangle,
		\end{equation}
		where the additional generators $\{F^{(i)}\}_{i=1}^{lq}$ are given by the entrywise decomposition
		$$
		\{F^{(i)}\}_{i=1}^{lq} = \mathcal{E}(F),
		$$
		with
		\begin{equation} \label{eq:Fgen}
			F \;=\; {\Delta}\!\left( \,\lvert M_C \rvert + \sum_{j=1}^g \lvert G^{(j)} \rvert \,\right) \;\in\; \mathbb{R}^{l \times q}.
		\end{equation}
	\end{definition}
	\begin{proposition}\label{thm:overapprox}
		Let $\II \in \mathbb{R}^{l \times p}$ be an interval matrix and $\MM\subseteq \mathbb{R}^{p \times q}$ be a matrix zonotope, then $\II\MM\subseteq\TT_{\II}(\MM).$
	\end{proposition}
	\emph{Proof. } See Appendix~\ref{app:overapprox}

	\begin{definition}\label{def:jfold}
		For $j \in \mathbb{N}$, we define the $j$-fold application of the operator $\TT_{\II}$ as
		$$
		\TT_{\II}^j(\MM) \;=\; 
		\underbrace{\TT_{\II} \circ \TT_{\II} \circ \cdots \circ \TT_{\II}}_{j \text{ times}} (\MM) 
		$$
	\end{definition}
	By iteratively applying Theorem~\ref{thm:overapprox}, one has 
	\begin{equation}
		\label{eq:IjMapprox}
		\II^j\MM\subseteq\TT_{\II}^j(\MM).
	\end{equation}

	\section{Problem Formulation}\label{sec:PF}
	Consider the linear time-invariant system
	\begin{equation}\label{eq:sys}
		x(k+1) = Ax(k) + Bu(k),
	\end{equation}
	where $x(k) \in\RR^n$ is the system state at time $k$, $u(k) \in\RR^m$ is the control input. The system matrices $A$ and $B$ are unknown and they belong to an interval matrix set, i.e., 
	\begin{equation}\label{eq:uncertainty}
		\interval{A\quad B}\in\MD =\interval{\Ac\quad\Bc}\oplus \intsimple{\Delta_A\quad \Delta_B},
	\end{equation}
	with $\Ac,\,\Bc,\,\Delta_A$ and $\Delta_B$ known. Let us denote $\Mdelta=\intsimple{\Delta_S}$ with $\Delta_S=\left[\Delta_A\quad\Delta_B\right]$. Then, system~\eqref{eq:sys} can be rewritten as $$x(k+1) = \Ac x(k) + \Bc u(k) + A_{\Delta} x(k) + B_{\Delta} u(k),$$
	where $A_{\Delta} = A-\Ac$, $B_{\Delta} = B-\Bc$ and $\left[A_{\Delta}\quad B_{\Delta}\right]\in \Mdelta$.
	The control objective is to steer the system state to the origin in minimum time, despite the presence of uncertainty. Moreover, the state and input are subject to the constraints
	\begin{equation}\label{eq:constraint}
		x(k) \in \XX(k),\,u(k) \in \UU(k),
	\end{equation}
	with $\XX(k),\,\UU(k)$ being convex polytopic sets.
	
	Consider now the nominal system, given by 
	\begin{equation}\label{eq:sys_nominal}
		z(k+1) = \Ac z(k) + \Bc v(k).
	\end{equation}
	Let the feedback policy for system~\eqref{eq:sys} be designed as
	\begin{equation}\label{eq:u_RMPC}
		u(k) = K(x(k)-z(k)) + v(k),
	\end{equation} 
	where $v(k)$ is the nominal control input and $K$ is such that the nominal closed-loop matrix $\AKc = \Ac+\Bc K$ is Schur. Clearly, the true closed-loop matrix $A_K=A+BK$ is uncertain and one has 
	\begin{equation}\label{eq:AK}
		A_K\in\AK = \AKc \oplus \Mdelta\myvec{I}{K}. 
	\end{equation}
	{The error between the true and the nominal state is defined as $e(k) = x(k)-z(k)$ and its dynamics evolves as
		\begin{equation}\label{eq:error_dyn2}
			e(k+1) = A_K e(k) + A_{\Delta} z(k) + B_{\Delta} v(k).
		\end{equation}
		Therefore, due to the uncertainty affecting $A$ and $B$, the error $e(k)$ evolves within a set sequence $\SS(k)$, i.e. $e(k)\in\SS(k)$, where $\SS(k)$ satisfy the set-valued dynamics
		\begin{equation}\label{eq:Sk_recursion}
			\SS(k+1) = \AK\SS(k) \oplus \Mdelta\left[\begin{array}{c}
				z(k)\\
				v(k)
			\end{array}\right].
	\end{equation}}

	In this work, the aim is to compute the nominal input $v(k)$ by solving a time-optimal control problem defined as follows
	\begin{equation}\label{eq:mpc_tube_exact} 
		\begin{aligned}
			\underset{N_k,\textbf{v}(k),\textbf{z}(k)}{\text{min }} &N_k\\
			\text{s.t.} \quad & z_k(0)= x(k)\\
			& {z_k}(j+1)=\Ac {z_k}(j)+\Bc {v_k}(j)  \\
			&  {z_k}(j)\oplus\SS_k(j) \subseteq \XX(k\!+\!j), \quad j=0,\ldots,N_k\!-\!1 \\
			&  v_k(j)\oplus K\SS_k(j)\subseteq \UU(k\!+\!j), \quad j=0,\ldots, N_k\!-\!1 \\
			& z_k(N_k) = 0\\
			& N_k \in \mathbb{N}^+
		\end{aligned}
	\end{equation}
	where $\mathbf{v}_k= \left[v_k(0),\ldots,\,v_k(N_k-1)\right]$ and $\mathbf{z}_k = \left[z_k(0),\ldots,\,z_k(N_k)\right]$ are the nominal input and state sequences, respectively, along the prediction horizon of length $N_k$. Notice that $N_k$ itself is an optimization variable in~\eqref{eq:mpc_tube_box}.  {The control input applied to the system at each time step $k$ is then given by $u(k) = v_k^*(0)$, according to the standard MPC implementation.}
	
	In accordance with~\eqref{eq:error_dyn2}, the error sets in~\eqref{eq:mpc_tube_exact} are defined as $$\SS_k(j+1) = \AK\SS_k(j)\oplus\Mdelta\left[\begin{array}{c}
		z_k(j)\\
		v_k(j) 
	\end{array}\right].$$  
	with the initial set being $\SS_k(0) = \{0\}$, due to the initial constraint $z_k(0) = x(k)$.
	The sets $\SS_k(j)$ can be rewritten as
	\begin{equation}\label{eq:Sk_response}
		\SS_k(j) =  \sum_{i=0}^{j-1}\AK^{j-i-1}\Mdelta\left[\begin{array}{c}
			z_k(i)\\
			v_k(i)
		\end{array}\right].
	\end{equation}
	
	Solving problem~\eqref{eq:mpc_tube_exact} is practically intractable for two reasons. 
	First, the computation of the exact error sets $\SS_k(j)$ is prohibitive. This is due both to the need for multiple exact set multiplications in~\eqref{eq:Sk_response}, and to their dependence on the optimization variables $\mathbf{v}_k$ and $\mathbf{z}_k$, which prevents any offline pre-computation. 
	Moreover, the set inclusions involved in the state and input constraints are highly impractical for standard solvers, since in general they require the computation of the vertices of polytopic sets. 
	Second, recursive feasibility of problem~\eqref{eq:mpc_tube_exact} is not guaranteed, due to the multiplicative uncertainty, affecting the true system dynamics.
	To address these issues, the next section introduces a tractable reformulation of problem~\eqref{eq:mpc_tube_exact}, which enables efficient online computation, along with a control strategy designed to ensure recursive feasibility and finite-time convergence of the overall control scheme.
	
	\section{Time-Optimal Robust MPC}\label{sec:TOR}
	The main idea of the tractable reformulation is to replace the error sets $\SS_k(j)$ appearing in the state and input constraints of problem~\eqref{eq:mpc_tube_exact} with suitable interval-shaped sets that can be computed offline. Moreover, a suitable terminal set is introduced in order to guarantee recursive feasibility. Therefore, the control scheme~\eqref{eq:mpc_tube_exact}, is modified in the new problem $\Pk$ defined as follows.
	\begin{subequations}\label{eq:mpc_tube_box}
		\begin{align}
			\min_{N_k,\mathbf v(k),\mathbf z(k)}\ & N_k \notag\\
			\text{s.t.}\quad & z_k(0)= x(k) \label{subeq:mpc_init}\\
			& z_k(j+1)=\Ac z_k(j)+\Bc v_k(j) \label{subeq:mpc_dynamic}\\
			& z_k(j)\oplus\BB_k(j) \subseteq \XX(k\!+\!j), \quad j=0,\ldots,N_k\!-\!1  \label{subeq:mpc_state}\\
			& v_k(j)\oplus K\BB_k(j)\subseteq \UU(k\!+\!j), \quad \!\!j=0,\ldots,\! N_k\!-\!1 \label{subeq:mpc_input}\\
			& z_k(N_k) \in \ZZ_{f,k} \label{subeq:mpc_terminal}\\
			& N_k \in \mathbb{N}^+, \notag
		\end{align}
	\end{subequations}
    where $\ZZ_{f,k}$ is a suitable terminal set, whose design is discussed below.
	The sets $\BB_k(j)$ in~\eqref{subeq:mpc_state}-\eqref{subeq:mpc_input} are defined as 
	\begin{equation}\label{eq:Bbox}
		\BB_k(j) = \sum_{i=0}^{j-1} \II(j-i-1)\myvec{z_k(i)}{v_k(i)},
	\end{equation}
	where
	\begin{equation}\label{eq:Ibox}
		\II(j) =  \bbox{\TT^j_{\MDK}(\MM_{\Delta})}
	\end{equation}
	with $\MM_{\Delta} = \langle 0;\,\mathcal{E}(\Delta_S)\rangle$ being the zonotopic representation of $\Mdelta$, and $$\MDK = \AKc \oplus \intsimple{\Delta_K},$$
	where $\Delta_K=\Delta_A + \Delta_B |K|$. 
	From \eqref{eq:intxM} and~\eqref{eq:AK} one has that $\AK  \subseteq \MDK$. Moreover, as a consequence of \eqref{eq:IjMapprox} one gets $\AK^{j}\Mdelta \subseteq \II(j)$. Therefore, from \eqref{eq:Sk_response} and \eqref{eq:Bbox}, it can be concluded that $\SS_k(j) \subseteq \BB_k(j),\,\forall k,j$. This means that constraints \eqref{subeq:mpc_state}-\eqref{subeq:mpc_input} are a conservative version of the corresponding constraints in problem \eqref{eq:mpc_tube_exact}.
	
	\begin{remark}
		{It is worth stressing that the role of the operator $\TT^j$ is crucial in reducing the conservatism in the outer bounding of the sets $\SS_k(j)$. In fact, it can be observed that the interval matrix that bounds $\AK^j \II_\Delta$, obtained by using $j$ consecutive matrix zonotope approximations as in \eqref{eq:Ibox}, is generally much less conservative than that resulting from $j$ consecutive interval matrix products, i.e., $\MDK * \MDK * \ldots * \MDK * \Mdelta$.
			Moreover, the use of interval matrices in~\eqref{eq:mpc_tube_box} allows one to treat the dependence on the optimization variables $\mathbf{z}_k$ and $\mathbf{v}_k$ in an efficient way, as it will be explained in Section~\ref{subsec:implementation}.}
	\end{remark}
	
	\subsection{Adaptive Terminal Set Design for Recursive Feasibility}
	To guarantee feasibility of the control action at each time step $k$ and finite-time convergence, the control mechanism adopted in this work builds upon the adaptive terminal set strategy introduced in~\cite{quartullo2025robust} for the case of additive disturbances. In particular, the key idea is to keep enforcing the terminal equality constraint $z_k(N_k) = 0$ (i.e., setting $\ZZ_{f,k} = \{0\}$) whenever the optimal horizon length decreases by at least 1 at each step.  {If this is not possible due to system uncertainty, the terminal set is enlarged in a way that preserves recursive feasibility, by taking into account the effect of multiplicative uncertainties in system~\eqref{eq:sys}. More specifically, whenever the optimal horizon length does not decreases at time step $k$, the terminal set is modified as
		\begin{equation}
			\ZZ_{f,k} = \ZZ_{f,k-1} \oplus \AKc^{N^*_{k-1}-1}\Mdelta\left[\begin{array}{c}
				x(k-1)\\
				u(k-1)
			\end{array}\right].
		\end{equation}
		This is instrumental to make problem $\Pk$ feasible and to force the optimal horizon length to decrease.
	}
	The overall control strategy is detailed in Algorithm~\ref{alg:control}.
	
	\begin{algorithm}[h]
		\caption{ {Time-Optimal Robust MPC (TOR-MPC)}}\label{alg:control}
		\begin{algorithmic}[1]
			\State \textbf{Input} $x(0)$
			\State Solve $\mathbb{P}_0(x(0),\{0\})$ and get $(N_0^*,\,\textbf{v}_0^*,\,\textbf{z}_0^*)$
			\State {$\ZZ_{f,0} \gets \{0\}$}
			\State $u(0) \gets v^*_0(0)$
			\State $x(1) \gets Ax(0)+Bu(0)$
			\State $k \gets 0$, $T_l \gets 0$
			\While{$N_k^*>1$} 
			\State $k \gets k+1$
			\State Solve $\mathbb{P}_k(x(k),\{0\})$, get $(\tilde{N}_k^*,\tilde{\textbf{v}}_k^*,\tilde{\textbf{z}}_k^*)$ 
			\If{ $\tilde{N}_k^* > N^*_{k-1} - 1$}
			\State $\ZZ_{f,k} \gets \ZZ_{f,k-1} \oplus \AKc^{N^*_{k-1}-1}\Mdelta\left[\begin{array}{c}
				x(k-1)\\
				u(k-1)
			\end{array}\right]$
			\State   Solve $\Pk$ 
			and get $({N}_k^*,\textbf{v}_k^*,\textbf{z}_k^*)$ 
			\Else  
			\State $({N}_k^*,\textbf{v}_k^*,\textbf{z}_k^*) \gets ( \tilde{N}_k^*,\tilde{\textbf{v}}_k^*,\tilde{\textbf{z}}_k^*)$
			\State  $\ZZ_{f,k} \gets \{0\}$
			\State $T_l\gets k$
			\EndIf
			\State $u(k) \gets v^*_k(0)$
			\State $x({k+1}) \gets Ax(k)+Bu(k)$
			\EndWhile
			\State \Return $T_l$, $T_c \gets k+1$
		\end{algorithmic}
	\end{algorithm}
	
	To establish robust finite-time convergence to a suitable neighborhood of the origin, it is useful to introduce the notion of \emph{shrinking recursive feasibility}.
	\begin{definition}
		The MPC control scheme arising from~\eqref{eq:mpc_tube_box} is said to be \emph{shrinking recursively feasible} if problem $\Pk$ is feasible at time $k=0$ and, for every $k \geq 1$, admits a feasible solution of length $N_{k-1}^* - 1$.
	\end{definition}
	The following results formalize feasibility and convergence properties of Algorithm~\ref{alg:control}.
	{\begin{theorem}\label{thm:RF}
			Let problem $\mathbb{P}_0(x(0),\{0\})$ be feasible. Then, the control strategy proposed in Algorithm~\ref{alg:control} is shrinking recursively feasible.
		\end{theorem}
		\emph{Proof.} See Appendix~\ref{app:RF}.}
	\begin{theorem}\label{thm:convergence}
		The closed-loop trajectories $x(k)$ of the uncertain system~\eqref{eq:sys}-\eqref{eq:uncertainty}, with the time-optimal MPC control law defined in Algorithm~\ref{alg:control}, converge in a finite number of steps $T_c$ to the set 
		\begin{equation}\label{eq:terminal_set}
			\mathcal{T} = \sum_{k = T_l}^{T_c-1} \AKc^{N^*_k-1}\Mdelta\myvec{x(k)}{u(k)}
		\end{equation}
		with $T_l$ as returned by Algorithm~\ref{alg:control}. 
		Moreover, $T_c \leq  N_0^*$ where $ N_0^* $ is the optimal horizon length of the initial problem $\mathbb{P}_0(x(0),\{0\})$. 
	\end{theorem}
	\emph{Proof.} See Appendix~\ref{app:convergence}.
	{\begin{remark}\label{remark:K}
			Notice that for Theorems~\ref{thm:RF} and~\ref{thm:convergence}, it is not necessary to assume that the matrix gain $K$ stabilizes system~\eqref{eq:sys} for all possible $\left[A\quad B\right]\in\MD$. However, requiring all possible $A_K$ to be Schur provides a convenient way to limit the growth of the error sets $\BB(k)$ during propagation and therefore allows one to enlarge the set of initial states $x(0)$ for which problem $\mathbb{P}_0(x(0),\{0\})$ admits a feasible solution.
	\end{remark}}
	
	{\subsection{Computational aspects}\label{subsec:implementation}}
	Recall that state constraints are polytopic sets, i.e., $\XX(j) = \{x\in\RR^n|\,\,H(j)x\leq b(j)\}$ and all interval matrices $\II(j) = \intsimple{\Delta_I(j)}$  are symmetric (i.e., the center is zero).
	Hence, the $j$-th state constraint~\eqref{subeq:mpc_state}, rewritten as
	$$z_k(j) \oplus \sum_{i=0}^{j-1}\II(j-1-i)\myvec{z_k(i)}{v_k(i)}\subseteq \XX(k+j),$$ 
	can be equivalently enforced as
	\begin{equation}\label{eq:constr_efficient}
		\begin{aligned}
			H(k+j)z_k(j) 
			&+ \left|H(k+j)\right|
			\sum_{i=0}^{j-1} \Delta_I(j-i-1)
			\left|\myvec{z_k(i)}{v_k(i)}\right| \\
			&\leq b(k+j).
		\end{aligned}
	\end{equation}
	The input constraint~\eqref{subeq:mpc_input} can be treated similarly. The constraint~\eqref{eq:constr_efficient} can be easily cast as a linear constraint (see, e.g.,~\cite{boyd2004convex}) and thus it can be efficiently handled online by standard solvers. This efficiency is enabled by the use of interval matrices in the optimization problem~\eqref{eq:mpc_tube_box} to bound the system error. Problem~\eqref{eq:mpc_tube_box} can be formulated as a mixed-integer linear program (MILP). Hence, efficient heuristic approaches can be adopted to speedup computations, as those presented in~\cite{leomanni2022variable,persson2024optimization}.
	\begin{remark}\label{remark:vertices}
		The proposed set bounding approach eliminates the need for vertex evaluation of the uncertainty set $\MD$ during online computations, unlike the approaches in~\cite[Ch.~5]{CANNONbook}, which become intractable for high-dimensional systems. In addition, vertex-based propagation of the error set $\SS(k)$ is also avoided.
	\end{remark}

	\section{Numerical example}\label{sec:simulations}
	The effectiveness of the proposed control law is demonstrated through a numerical case study involving an orbital rendezvous mission between a controlled (chaser) and an uncontrolled (target) satellite~\cite{shan2016review,leomanni2020sum}. The evolution of relative position and velocity, expressed in the radial-transverse-normal (RTN) reference frame, evolves according to the Hill-Clohessy-Wiltshire (HCW) model~\cite{clohessy1960terminal}. For the MPC implementation, the HCW equations are  discretized with a sampling time of 11.7~s. Considering a nominal mean motion of $0.001$~rad/s, the resulting nominal system matrices are
	\begin{equation*}
		\Ac = \begin{bmatrix}
			1      & 0       & 0       & 11.7       & 0       & 0       \\
			0       & 1       & 0       & 0       & 11.7       & 0       \\
			0       & 0       & 1      & 0       & 0       & 11.7       \\
			3.8\cdot 10^{-5}  & 0       & 0       & 1       & 0.02  & 0       \\
			0       & 0       & 0       & -0.02 & 1       & 0       \\
			0       & 0       & -1.3\cdot 10^{-5} & 0       & 0       & 1
		\end{bmatrix}
	\end{equation*}
	\begin{equation*}
		\Bc = 11.7\begin{bmatrix}
			0       & 0       & 0            \\
			0       & 0       & 0           \\
			0       & 0       & 0              \\
			1  & 0       & 0             \\
			0      & 1       & 0             \\
			0      & 0       &  1 
		\end{bmatrix}.
	\end{equation*}
	The control input $u(k)$ represents the acceleration along the RTN axes. In this study, the propulsion system is assumed to be affected by imperfect mounting, leading to a possible thrust misalignment along each axis of up to 1~deg. We also assume that the some entries of the matrix $A$ are uncertain due to a 5\% uncertainty in the mean motion value. Hence, the interval matrix $\intsimple{\Delta_A\,\,\Delta_B}$ in~\eqref{eq:uncertainty} is defined as
	\begin{equation*}
		\Delta_A = 10^{-3}\begin{bmatrix}
			0       & 0       & 0       & 0       & 0       & 0       \\
			0       & 0       & 0       & 0       & 0       & 0       \\
			0       & 0       & 0       & 0       & 0       & 0       \\
			0.004  & 0       & 0       & 0       & 1.23  & 0       \\
			0       & 0       & 0       & 1.23 & 0       & 0       \\
			0       & 0       & 0.001 & 0       & 0       & 0
		\end{bmatrix},
	\end{equation*}
	\begin{equation*}
		\Delta_B = 0.205\begin{bmatrix}
			0       & 0       & 0            \\
			0       & 0       & 0           \\
			0       & 0       & 0              \\
			0  & 1       & 1             \\
			1       & 0       & 1             \\
			1      & 1       &  0 
		\end{bmatrix}.
	\end{equation*}
	In order to maintain visual contact with the target, the relative position is constrained to lie within a visibility cone characterized by a view angle of 60~deg. This cone is then approximated by an inner polyhedral set. The relative maximum velocity is 0.4~{m/s} along each direction, while the propulsion system is limited to a maximum acceleration of $0.01~\text{m/s}^2$. These specifications define the polytopic state and input constraints in~\eqref{eq:constraint}. The feedback gain matrix is chosen as 
	$$\!K \!= \!-0.1\!\begin{bmatrix}
		0.025 & 0       & 0       & 1.005 & 0.021 & 0       \\
		0       & 0.026 & 0       & -0.021  & 1.022 & 0       \\
		0       & 0       & 0.026 & 0       & 0       & 1.022
	\end{bmatrix}$$ 
	ensuring that all admissible closed-loop matrices $A_K$ are Schur stable (see Remark~\ref{remark:K}).
	\begin{figure}
		\centering
		\includegraphics[width=0.7\linewidth]{ 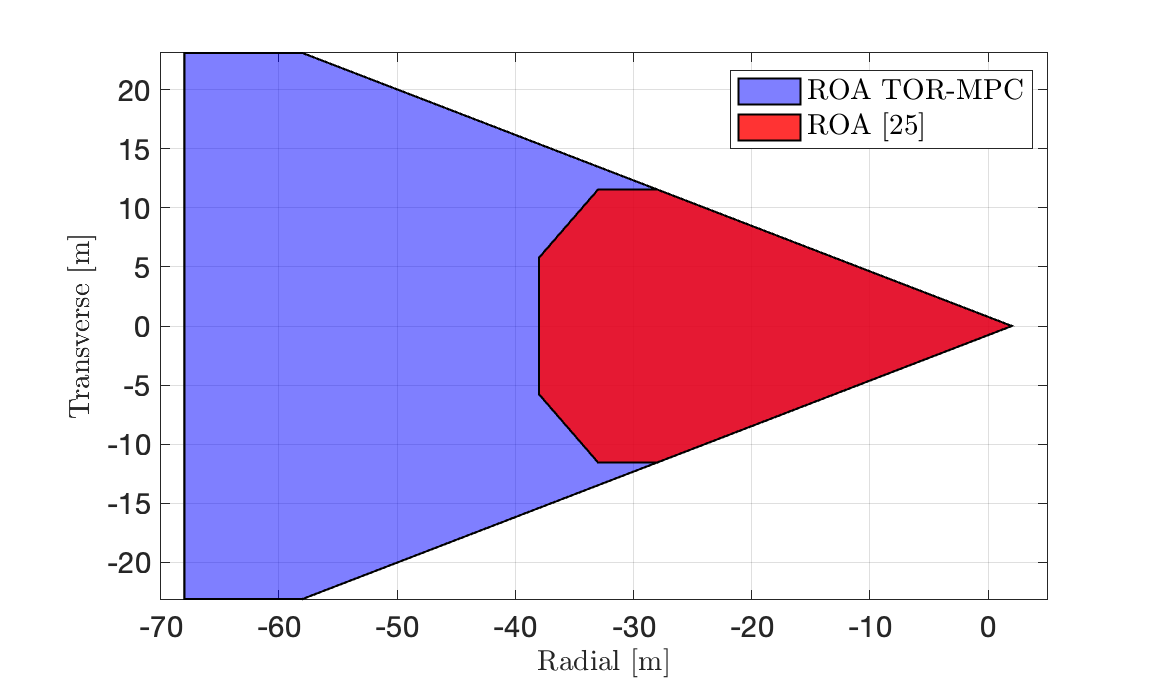}
		\caption{ROA comparison between TOR-MPC (blue) and~\cite{quartullo2024robust} (red).}
		\label{fig:ROA}
	\end{figure}
	\begin{figure}
		\centering
		\includegraphics[width=0.7\linewidth]{ 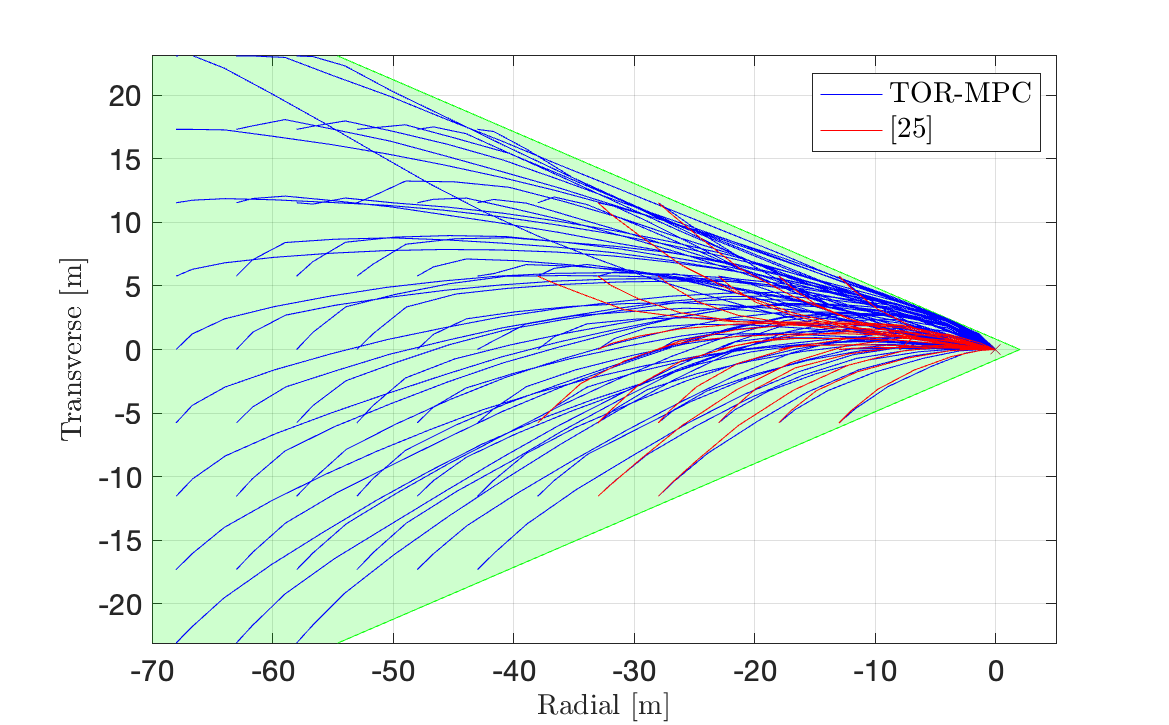}
		\caption{Relative position trajectories obtained by applying TOR-MPC (blue) and~\cite{quartullo2024robust} (red). The green set is the visibility region $\XX$.}
		\label{fig:traj}
	\end{figure}
	
	{The proposed Time-Optimal Robust MPC (TOR-MPC) is compared to the control law introduced in~\cite{quartullo2024robust}, in which the uncertain system~\eqref{eq:sys} is represented as $x(k+1) = \Ac x(k) + \Bc u(k) + w(k)$, with $w(k)\in \WW $. The set $\WW$ is defined to over-bound the uncertainty term $\delta(k) =A_{\Delta}x(k) + B_{\Delta}u(k)\in \Mdelta\myvec{x(k)}{u(k)}$ for all possible values of $x(k)$ and $u(k)$ satisfying constraints~\eqref{eq:constraint}.}
	
	A set of 75 initial conditions are considered. These are selected on the same orbital plane as the target so as to uniformly cover the entire visibility region, with a maximum radial separation of 70~m. The initial relative velocity is set to zero. In each simulation, the matrices $A_{\Delta}$ and $B_{\Delta}$ are randomly sampled from $\intsimple{\Delta_A}$ and  $\intsimple{\Delta_B}$, respectively. 
	
	Figure~\ref{fig:ROA} illustrates the convex hull of the initial conditions for which problem $\mathbb{P}_0(x(0),{0})$ is feasible. This convex hull approximates the region of attraction (ROA) corresponding to each control scheme. It turns out that the proposed control law is feasible for all initial conditions within the visibility region. In contrast, the ROA associated with the approach in~\cite{quartullo2024robust} is significantly smaller, with only about 32\% of the tested initial conditions lying within it. This behavior is clearly a consequence of the conservative over-approximation of the multiplicative uncertainty when described as an additive disturbance. The reduced conservatism of the proposed control law is also evident in Fig.~\ref{fig:traj}, which shows the relative position trajectories. It can be observed that the trajectories generated by the law in~\cite{quartullo2024robust} remain farther from the boundaries of $\XX$. As a consequence, the control scheme in~\cite{quartullo2024robust} exhibits an 18\% higher fuel consumption.
	\begin{figure}[t]
		\centering
		\begin{subfigure}[t]{0.48\columnwidth}
			\centering
			\includegraphics[width=\linewidth]{ 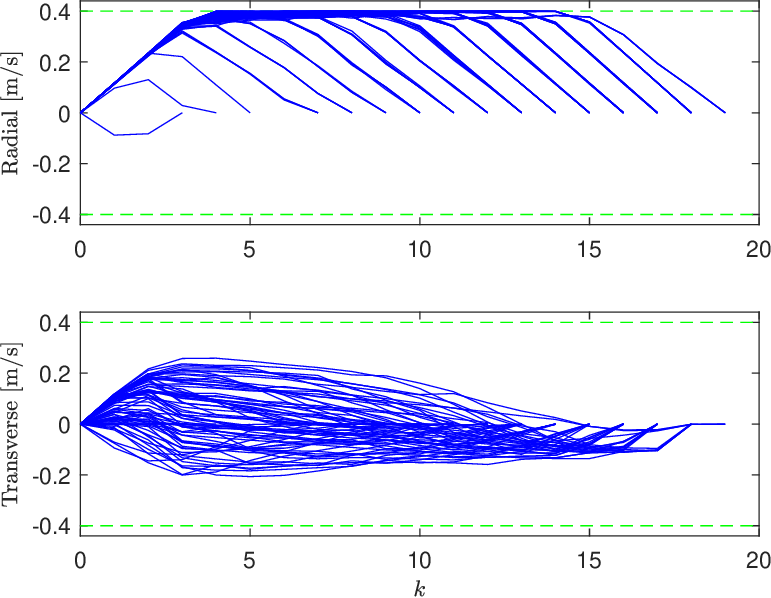}
			\caption{}
			\label{fig:v}
		\end{subfigure}
		\hfill
		\begin{subfigure}[t]{0.48\columnwidth}
			\centering
			\includegraphics[width=\linewidth]{ 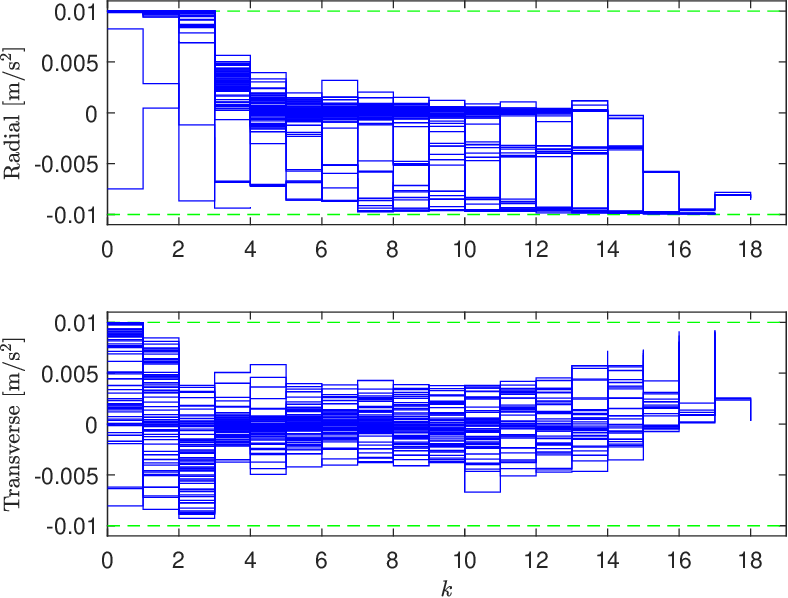}
			\caption{}
			\label{fig:u}
		\end{subfigure}
		\caption{\new{Relative velocity (a) and input acceleration (b) profiles obtained by applying TOR-MPC in all tested initial conditions.}}
	\end{figure}
	
	Figures~\ref{fig:v} and \ref{fig:u} show the relative velocity and input acceleration profiles resulting from the proposed TOR-MPC. It can be seen that the constraints on both relative velocity and acceleration (dashed green lines) are robustly satisfied throughout the maneuver. {Moreover, the average radius of the projection of the final set $\mathcal{T}$ onto the position subspace is 0.038~m, while its projection onto the velocity subspace has an average radius of 0.004~m/s. Consistently with Theorem~\ref{thm:convergence}, the average values of the final position and velocity errors of the resulting closed-loop trajectories are 0.016~m and 0.001~m/s, respectively.}
	
	All simulations are performed in Matlab on a standard laptop equipped with an Intel i7 processor and 16~GB of RAM. It is worth noting that the time require to solve problem~\eqref{eq:mpc_tube_box} is always less than 1~s at every time step. This computational time is well within the chosen sampling period, demonstrating the online tractability of the proposed scheme, even for a nontrivial case study involving a dynamic system with 6 states.

	\section{Conclusions}\label{sec:conclusions}
	A time-optimal robust MPC scheme for linear systems with interval uncertainties has been presented. The method exploits matrix-zonotope approximations to enable a computationally tractable uncertainty propagation. Recursive feasibility and finite-time convergence are ensured using an adaptive terminal set design. Numerical results on an orbital rendezvous example demonstrate reduced conservatism and real-time implementability on standard hardware. Future developments concern the extension of the proposed approach to variable-horizon MPC schemes, optimizing more general cost functions.
	
	\appendix
	\section{Appendix}
	\subsection{Proof of Proposition~\ref{thm:overapprox}}\label{app:overapprox}
	Let $\II$ and $\MM$ be as in Definition \ref{def:operatorT}.
	Then,
	\begin{equation}
		\label{eq:IM}
		\II \MM = (C \oplus \intsimple{\Delta}) \langle M_C;\, G^{(1)}, \ldots, G^{(e)}\rangle
		\subseteq \langle C M_C;\, C G^{(1)}, \ldots, C G^{(g)} \rangle \oplus \intsimple{\Delta}\MM.
	\end{equation}
	Since $\MM$ can be expressed as 
	$$\MM = M_C \oplus \langle 0;\, G^{(1)}, \ldots, G^{(g)} \rangle,$$ one has
	\begin{equation*}
		\begin{split}
			\intsimple{\Delta}\MM & \subseteq \intsimple{\Delta}M_C \oplus\intsimple{\Delta}\langle0;\,G^{(1)},\ldots,G^{(g)}\rangle\\
			& \subseteq \intsimple{\Delta|M_C|} \oplus \left\{ M:\,\,M \in \sum_{i=1}^g \intsimple{\Delta}G^{(i)}\intsimple{1}\right\}\\
			& \subseteq \intsimple{\Delta|M_C|} \oplus\sum_{i=1}^g  \intsimple{\Delta |G^{(i)}|}\\
			& = \intsimple{\Delta\left(|M_C| + \sum_{i=1}^g |G^{(i)}|\right)}=\intsimple{F}= \langle 0; \EE(F) \rangle.
		\end{split}
	\end{equation*}
	Hence, the results follows from \eqref{eq:IM}, by using the definition of Minkowski sum and noticing that $$ \langle C M_C;\, C G^{(1)}, \ldots, C G^{(g)} \rangle \oplus  \langle 0; \EE(F) \rangle$$ is equal to the definition of $\TT_{\II}(\MM)$ in \eqref{eq:Topdef}. \qed
	
	\subsection{Auxiliary lemmas}
	{
		In order to prove the main results in the paper, the following preliminary lemmas are needed.
		\begin{lemma}
			\label{lem:Ij}
			The set $\II(j)$ defined in \eqref{eq:Ibox} satisfies
			\begin{equation}
				\label{eq:Ijint}
				\II(j)= \intsimple{ \sum_{i=0}^j | \hat{A}_K^{j-i} | F_i }
			\end{equation}
			where $F_0=\Delta_S$ and
			\begin{equation}
				\label{eq:Fj}
				F_{j+1} = \Delta_K \sum_{i=0}^{j} | \hat{A}_K^{j-i} | F_i, ~~~\mbox{for}~j \geq 0. 
			\end{equation}
		\end{lemma}
		\emph{Proof. } Let us first prove by induction that 
		\begin{equation}
			\label{eq:TTj}
			\TT^j_{\MDK}(\MM_{\Delta})=\langle 0; \hat{A}_K^j \EE(F_0),   \hat{A}_K^{j-1} \EE(F_1), \dots,  \EE(F_j) \rangle.
		\end{equation}
		For $j=1$, by applying Definition \ref{def:operatorT} with $\II=\MDK=\hat{A}_K \oplus \intsimple{\Delta_K}$ and $\MM=\MM_{\Delta} = \langle 0;\,\mathcal{E}(\Delta_S)\rangle$ one gets
		$$
		\TT_{\MDK}(\MM_{\Delta})=\langle 0; \hat{A}_K \EE(F_0),  \EE(F_1) \rangle,
		$$
		with $F_0=\Delta_S$ and $F_1=\Delta_K \Delta_S$, which is in agreement with \eqref{eq:Fj}. Now, let \eqref{eq:TTj} hold for a generic $j$. By applying again Definition \ref{def:operatorT}  one obtains
		$$
		\TT^{j+1}_{\MDK}(\MM_{\Delta}) = \TT_{\MDK} \circ \TT_{\MDK}^j (\MM_{\Delta}) = 
		\langle 0; \hat{A}_K^{j+1} \EE(F_0),   \hat{A}_K^{j} \EE(F_1), \dots,  \hat{A}_K \EE(F_{j}), \EE(F_{j+1}) \rangle,
		$$
		{where, using \eqref{eq:Fgen} and \eqref{eq:AdecM}, one has
			$$
			\begin{array}{rcl}
				F_{j+1} &=& \displaystyle{ \Delta_K \sum_{i=1}^{n(n+m)} \left(  |\hat{A}_K^j F_0^{(i)}|+  |\hat{A}_K^{j-1} F_1^{(i)}|+ \dots + | F_j^{(i)} | \right) }\\ 
				&=& \Delta_K \left( |\hat{A}_K^j| F_0 + |\hat{A}_K^{j-1} | F_1 + \dots +  F_j \right)
			\end{array}
			$$
		}
		which corresponds to  \eqref{eq:Fj}.
		Then, \eqref{eq:Ijint} immediately follows by applying Proposition \ref{prop:box} to the right hand side of \eqref{eq:Ibox}.
		\begin{lemma}
			\label{lem:Pj}
			Let us define matrices $P_j$ such that $P_1=I$ and
			\begin{equation}
				\label{eq:Pj}
				P_{j+1} = | \hat{A}_K^j | + \sum_{h=1}^{j} P_h \Delta_K | \hat{A}_K^{j-h} |, ~~~\mbox{for}~j \geq 1. 
			\end{equation}
			Then, matrices $F_j$ in \eqref{eq:Fj} satisfy
			\begin{equation}
				\label{eq:FjPj}
				F_{j} = \Delta_K P_j \Delta_S, ~~~\mbox{for}~j \geq 1. 
			\end{equation}
		\end{lemma}
		\emph{Proof. } By recursively expanding the sum in \eqref{eq:Fj}, one gets that $F_j=\Delta_K \Sigma_{j-1} \Delta_S$, where $\Sigma_j$ is the sum of all terms $| \hat{A}_K^{j_1} | \Delta_K^{j_2}$, such that $j_1+j_2=j$, with $0 \leq j_1 \leq j $, $0 \leq j_2 \leq j$.  Similarly, by recursively expanding the sum in \eqref{eq:Pj}, one gets $P_j=\Sigma_{j-1}$, thus giving \eqref{eq:FjPj}. 
	}

	\subsection{Proof of Theorem~\ref{thm:RF}}\label{app:RF}
	Let us consider the following candidate solution for step $k+1$ 
	\begin{equation}\label{eq:zv_candidate}
		\begin{array}{l}
			\hat{v}_{k+1}(j) =
			v^*_k(j+1) + K\AKc^j\delta(k), \quad j = 0,\ldots,N^*_k-2,\\
			\hat{z}_{k+1}(j) =
			z^*_k(j+1) + \AKc^j\delta(k), \quad j = 0,\ldots,N^*_k-1
		\end{array}
	\end{equation}
	with 
	{
		\begin{equation}
			\label{eq:deltak}
			\delta(k) = x(k+1)-\hat{A} x(k) - \hat{B} u(k) = A_{\Delta} x(k) + B_{\Delta} u(k)
		\end{equation}
		and $z_k^*(j)$, $v_k^*(j)$ are the optimal nominal states and inputs for problem~\eqref{eq:mpc_tube_box}. Note that the length of this solution is $N_k^*-1$.
		It is worth stressing that $\delta(k)$ is a known vector at time $k+1$ (when the state $x(k+1)$ is available), and therefore also the candidate solution in \eqref{eq:zv_candidate} is known. Moreover, one has that
		\begin{equation}
			\label{eq:deltakset}
			\delta(k)
			\in \Mdelta\left[\begin{array}{c}
				x(k)\\
				u(k)
			\end{array}\right] 
			= \intsimple{\Delta_S}  \myvec{z_k^*(0)}{v_k^*(0)}.
		\end{equation}
	}
	
	\emph{Initial constraint.} The initial state of the candidate solution~\eqref{eq:zv_candidate} satisfies
	\begin{equation*}
		\begin{split}
			\hat{z}_{k+1}(0) & = z^*_k(1)+\delta(k)\\ 
			&= \Ac z^*_k(0)+\Bc v^*_k(0)+A_{\Delta}x(k) + B_{\Delta}u(k)\\
			& = (\Ac + A_{\Delta})x(k) + (\Bc + B_{\Delta})u(k) = x(k+1).
		\end{split}
	\end{equation*}
	
	\emph{Dynamic constraint.} Given~\eqref{eq:zv_candidate}, we have that
	\begin{equation*}
		\begin{split}
			&\hat{z}_{k+1}(j+1)-\AKc^{j+1}\delta(k) = z^*_k(j+2)\\ 
			& = \Ac z_k^*(j+1) + \Bc v_k^*(j+1)\\
			&= \Ac\left(\hat{z}_{k+1}(j)-\AKc^j\delta(k)\right)+\Bc\left(\hat{v}_{k+1}(j)-K\AKc^j\delta(k)\right) \\
			&= \Ac\hat{z}_{k+1}(j)+\Bc \hat{v}_{k+1}(j) - \AKc^{j+1}\delta(k).
		\end{split}
	\end{equation*}
	Hence,  $\hat{z}_{k+1}(j+1) = \Ac \hat{z}_{k+1}(j)+\Bc \hat{v}_{k+1}(j)$ implying satisfaction of constraint~\eqref{subeq:mpc_dynamic}.
	
	\emph{State constraint.}
	{
		Let us define 
		\begin{equation}
			\label{eq:BBkstar}
			\BB_k^*(j) = \sum_{i=0}^{j-1} \II(j-i-1) \xi_k^*(i)
		\end{equation}
		and
		\begin{equation}
			\hat\BB_{k+1}(j) = \sum_{i=0}^{j-1} \II(j-i-1) \hat\xi_{k+1}(i)
		\end{equation}
		where
		$$
		\xi_k^*(i) = \myvec{z_k^*(i)}{v_k^*(i)}\,,~~~
		\hat\xi_{k+1}(i) = \myvec{\hat{z}_{k+1}(i)}{\hat{v}_{k+1}(i)}.
		$$
		In order to prove that the candidate solution \eqref{eq:zv_candidate} satisfies constraint~\eqref{subeq:mpc_state}, it is sufficient to show that
		\begin{equation}
			\label{eq:setincl}
			\hat{z}_{k+1}(j) \oplus \hat\BB_{k+1}(j) \subseteq z_k^*(j+1) \oplus \BB_k^*(j+1) 
		\end{equation}
		for $j=0,1,\dots,N_k^*-1$, because the right hand side of \eqref{eq:setincl} is included in $\XX(k+j+1)$. 
		By using the second equation in  \eqref{eq:zv_candidate}, one has that \eqref{eq:setincl} is equivalent to
		\begin{equation}
			\label{eq:setincl2}
			\hat{A}_{K}^j \delta(k) \oplus \hat\BB_{k+1}(j) \subseteq \BB_k^*(j+1). 
		\end{equation}
		Exploiting \eqref{eq:zv_candidate}, the left hand side of \eqref{eq:setincl2} can be rewritten as \begin{equation*}
			\begin{split}
				& \hat{A}_{K}^j \delta(k) \oplus \hat\BB_{k+1}(j) \\
				& = \hat{A}_{K}^j \delta(k) \oplus \sum_{i=0}^{j-1} \II(j-i-1) \hat\xi_{k+1}(i) \\
				& = \hat{A}_{K}^j \delta(k) \oplus \sum_{i=0}^{j-1} \II(j-i-1) \xi_{k}^*(i+1) \oplus \sum_{i=0}^{j-1} \II(j-i-1) M_K \hat{A}_K^i \delta(k) 
			\end{split}
		\end{equation*}
		where $M_K=\myvec{I}{K}$.
		Therefore, by using \eqref{eq:BBkstar}, \eqref{eq:setincl2} boils down to
		\begin{equation}	
			\label{eq:setincl3}
			\hat{A}_{K}^j \delta(k) \oplus \sum_{i=0}^{j-1} \II(j-i-1) M_K \hat{A}_K^i \delta(k)  \subseteq \II(j) \xi_k^*(0). 
		\end{equation}
		Now, using Lemmas \ref{lem:Ij} and \ref {lem:Pj} {and recalling that $\Delta_K = \Delta_S |M_K|$}, the left hand side of \eqref{eq:setincl3} satisfies 
		\begin{align}
			& \hat{A}_{K}^j \delta(k) \oplus \sum_{i=0}^{j-1} \II(j-i-1) M_K \hat{A}_K^i \delta(k)  \notag \\
			& \overset{\eqref{eq:Ijint}}{=} \hat{A}_{K}^j \delta(k) \oplus \sum_{i=0}^{j-1} \intsimple{ \sum_{h=0}^{j-1-i} | \hat{A}_K^{j-1-i-h} | F_h } M_K \hat{A}_K^i \delta(k)   \notag\\
			& \overset{\eqref{eq:FjPj}}{=} \hat{A}_{K}^j \delta(k) \oplus \sum_{i=0}^{j-1} \intsimple{ | \hat{A}_K^{j-1-i} | \Delta_S + \sum_{h=1}^{j-1-i} | \hat{A}_K^{j-1-i-h} | \Delta_K P_h \Delta_S } M_K \hat{A}_K^i \delta(k)  \notag \\
			& \overset{\eqref{eq:intxM}}{\subseteq} \hat{A}_{K}^j \delta(k) \oplus \sum_{i=0}^{j-1} \intsimple{ \left\{ | \hat{A}_K^{j-1-i} | \Delta_S + \sum_{h=1}^{j-1-i} | \hat{A}_K^{j-1-i-h} | \Delta_K P_h \Delta_S \right\}\, |M_K| \, |\hat{A}_K^i | } \delta(k)  \notag\\
			& =  \hat{A}_{K}^j \delta(k) \oplus \sum_{i=0}^{j-1} \intsimple{ | \hat{A}_K^{j-1-i} | \Delta_K \, |\hat{A}_K^i | } \delta(k)  \oplus \sum_{i=0}^{j-1} \intsimple{ \sum_{h=1}^{j-1-i} | \hat{A}_K^{j-1-i-h} | \Delta_K P_h \Delta_K  \, |\hat{A}_K^i | } \delta(k) \notag \\
			&= \hat{A}_{K}^j \delta(k) \oplus \sum_{i=0}^{j-1} \intsimple{ | \hat{A}_K^{j-1-i} | \Delta_K \, |\hat{A}_K^i | }  \delta(k) \oplus  \sum_{h=1}^{j-1}  \sum_{i=0}^{j-1-h} \intsimple{ | \hat{A}_K^{j-1-i-h} | \Delta_K P_h \Delta_K  \, |\hat{A}_K^i | } \delta(k)  \label{eq:lastlhs}
		\end{align}
		\new{where the last two equalities exploit property \eqref{eq:intsum}} and a reorganization of the sum indexes.
		Similarly, the right hand side of \eqref{eq:setincl3} satisfies
		\begin{align}
			& \II(j) \xi_k^*(0) =  \intsimple{ \sum_{i=0}^{j} | \hat{A}_K^{j-i} | F_i }  \xi_k^*(0) \notag \\
			& \overset{\eqref{eq:FjPj}}{=}  \intsimple{| \hat{A}_K^{j} | \Delta_S + \sum_{i=1}^{j} | \hat{A}_K^{j-i} |  \Delta_K P_i \Delta_S}  \xi_k^*(0)  \notag \\
			& \overset{\eqref{eq:Pj}}{=}  \intsimple{| \hat{A}_K^{j} | \Delta_S + \sum_{i=1}^{j} | \hat{A}_K^{j-i} |  \Delta_K  | \hat{A}_K^{i-1} | \Delta_S  + \sum_{i=1}^{j} | \hat{A}_K^{j-i} | \Delta_K  \sum_{h=1}^{i-1} P_h \Delta_K | \hat{A}_K^{i-1-h} |  \Delta_S }  \xi_k^*(0)  \notag\\
			& =  \intsimple{| \hat{A}_K^{j} | \Delta_S + \sum_{i=0}^{j-1} | \hat{A}_K^{j-1-i} |  \Delta_K  | \hat{A}_K^{i} | \Delta_S  + \sum_{i=0}^{j-1} | \hat{A}_K^{j-1-i} | \Delta_K  \sum_{h=1}^{i} P_h \Delta_K | \hat{A}_K^{i-h} |  \Delta_S }  \xi_k^*(0)  \notag \\
			& =  \intsimple{| \hat{A}_K^{j} | \Delta_S + \sum_{i=0}^{j-1} | \hat{A}_K^{j-1-i} |  \Delta_K  | \hat{A}_K^{i} | \Delta_S  + \sum_{h=1}^{j-1}  \sum_{i=h}^{j-1} | \hat{A}_K^{j-1-i} | \Delta_K  P_h \Delta_K | \hat{A}_K^{i-h} |  \Delta_S }  \xi_k^*(0)  \notag \\
			& = \intsimple{| \hat{A}_K^{j} | \Delta_S + \sum_{i=0}^{j-1} | \hat{A}_K^{j-1-i} |  \Delta_K  | \hat{A}_K^{i} | \Delta_S  + \sum_{h=1}^{j-1}  \sum_{i=0}^{j-1-h} | \hat{A}_K^{j-1-i-h} | \Delta_K  P_h \Delta_K | \hat{A}_K^{i} |  \Delta_S }  \xi_k^*(0).   \label{eq:lastrhs}
		\end{align}
		Now, recall that from \eqref{eq:deltakset} one has $\delta(k) \in \intsimple{\Delta_S}\xi_k^*(0)$. Therefore, by comparing \eqref{eq:lastlhs} and \eqref{eq:lastrhs}, and using properties \eqref{eq:intprod} and \eqref{eq:intxMleft}, one gets that \eqref{eq:setincl3} holds, and therefore also \eqref{eq:setincl}.
	}
	
	\emph{Input constraint.} The proof of input constraint~\eqref{subeq:mpc_input} satisfaction of solution~\eqref{eq:zv_candidate} follows the same arguments used for state constraint.
	
	\emph{Terminal constraint.} Being $z^*_k(N_k^*)\in\ZZ_{f,k}$, then one has
	\begin{equation*}
		\begin{split}
			\hat{z}_{k+1}(N_k^*\!-\!1) &= z_k^*(N_k^*) + \AKc^{N_k^*-1}\delta(k)  \\ 
			&\in\ZZ_{f,k} \oplus \AKc^{N_k^*-1}\Mdelta\myvec{x(k)}{u(k)}\\
			&  = \ZZ_{f,k+1}.
		\end{split}
	\end{equation*}
	\new{Since, the candidate solution~\eqref{eq:zv_candidate} is feasible for problem $\mathbb{P}_{k+1}(x(k),\ZZ_{f,k+1})$ and its associated horizon length is $N_k^*-1$, then the control strategy in Algorithm~\ref{alg:control} is shrinking recursively feasible.\qed}
	
	{\subsection{Proof of Theorem~\ref{thm:convergence}}\label{app:convergence}
		The upper bound on the convergence time $T_c$ is a direct consequence of the shrinking horizon property, ensured by Theorem~\ref{thm:RF}. Moreover, according to Algorithm~\ref{alg:control}, at step $T_c-1$, problem $\mathbb{P}_{T_c-1}(x(T_c-1),\ZZ_{f,T_c-1})$ is solved (for the last time) with the terminal set $\ZZ_{f,T_c-1} = \sum_{k = T_l}^{T_c-2} \AKc^{N^*_k-1}\Mdelta\myvec{x(k)}{u(k)}$, being $T_l$ the time step in which the control action is selected for the last time as the solution of problem $\mathbb{P}_k(x(k),\{0\})$. By definition of the termination condition in Algorithm~\ref{alg:control} (line 7), the resulting optimal horizon is $N^*_{T_c-1} = 1$. Hence, 
		\begin{equation*}
			\begin{split}
				x(T_c) &= \Ac x(T_c-1) + \Bc u(T_c -1) + \delta(T_c -1) \\
				&= z^*_{T_c-1}(1) + \delta(T_c -1)\\
				& \in \ZZ_{f,T_c-1} \oplus \Mdelta\myvec{x(T_c -1)}{u(T_c -1)} = \mathcal{T}
			\end{split}
		\end{equation*}
		with $\mathcal{T}$ defined in~\eqref{eq:terminal_set}.\qed}
	
	\bibliographystyle{ieeetr}
	\bibliography{biblioVH}

\end{document}